\documentclass{article}

\usepackage[english]{babel}
\usepackage{amssymb}
\usepackage{indentfirst}
\usepackage[letterpaper,top=2cm,bottom=2cm,left=3cm,right=3cm,marginparwidth=1.75cm]{geometry}

\usepackage{amsmath}
\usepackage{graphicx}
\usepackage{subfigure}
\usepackage[colorlinks=true, allcolors=blue]{hyperref}
\usepackage[numbers, sort&compress]{natbib}
\usepackage{caption}
\usepackage{authblk}
\title{Sharing tripartite nonlocality sequentially using only projective measurements}
\author[1,2]{Yiyang Xu}
\author[1,2]{Hao Sun}
\author[1,2]{Fenzhuo Guo\thanks{Corresponding author: gfenzhuo@bupt.edu.cn}}
\author[3]{Haifeng Dong}
\author[4]{Qiaoyan Wen}
\affil[1]{School of Science, Beijing University of Posts and Telecommunications, Beijing, 100876, China}
\affil[2]{Henan Key Laboratory  of  Network  Cryptography Technology, Zhengzhou, 450001, China}
\affil[3]{School of Instrumentation Science and Opto-Electronics Engineering, Beihang University, Beijing, 100191, China}
\affil[4]{State Key Laboratory of Networking and Switching Technology, Beijing University of Posts and Telecommunications, Beijing, 100876, China}

\begin{document}
\maketitle

\begin{abstract}
Bell nonlocality is a valuable resource in quantum information processing tasks. Scientists are interested in whether a single entangled state can generate a long sequence of nonlocal correlations. Previous work has accomplished sequential tripartite nonlocality sharing through unsharp measurements. In this paper, we investigate the sharing of tripartite nonlocality using only projective measurements and sharing classical randomness. For the generalized GHZ state, we have demonstrated that using unbiased measurement choices, two Charlies can share the standard tripartite nonlocality with a single Alice and a single Bob, while at most one Charlie can share the genuine tripartite nonlocality with a single Alice and a single Bob. However, with biased measurement choices, the number of Charlies sharing the genuine tripartite nonlocality can be increased to two. Nonetheless, we find that using biased measurements does not increase the number of sequential observers sharing the standard tripartite nonlocality. Moreover, we provide the feasible range of double violation for the parameters of the measurement combination probability with respect to the state.
\end{abstract}

\section{Introduction}

Quantum nonlocality is one of the most important properties of quantum mechanics. It was first pointed out by Einstein, Podolsky, and Rosen \cite{einstein1935can}, highlighting conflicts between quantum mechanics and local realism. Later, Bell derived a statistical inequality, known as the Bell inequality, using it to certify nonlocality \cite{bell1964einstein}. Subsequently, various Bell inequalities have been derived and extensively studied for nonlocality \cite{clauser1969proposed,svetlichny1987distinguishing,mermin1990extreme,ardehali1992bell,collins2002bell,brukner2002quantum,belinskiui1993interference,zukowski2002bell}, with experimental verifications conducted in many different scenarios \cite{aspect1982experimental,giustina2013bell,giustina2015significant,foletto2021experimental}. Moreover, Bell nonlocality \cite{brunner2014bell} serves as a valuable resource in quantum information processing tasks such as device independent randomness generation \cite{pironio2010random,acin2012randomness,acin2016certified,woodhead2018randomness}, quantum key distribution \cite{acin2007device}, and reductions of communication complexity \cite{buhrman2010nonlocality}.

The study of nonlocality sharing among multiple observers has been a hot topic. In 2015, Silva et. al. \cite{silva2015multiple} demonstrated through unsharp measurements that two Bobs could share the nonlocality with a single Alice. This opened up extensive research into the nonlocality sharing among multiple observers. In 2020, Brown and Colbeck considered the scenario where each Bob in the sequence performed unsharp measurements with unequal sharpness parameters \cite{brown2020arbitrarily}. They found that an arbitrary number of Bobs could share the nonlocality of a maximally entangled two-qubit state with a single Alice, and they extended this conclusion to all pure entangled two-qubit states. Zhang and Fei investigated sharing the nonlocality of arbitrary dimensional bipartite entangled \cite{zhang2021sharing}. In three-qubit system, Saha et. al. \cite{saha2019sharing} studied sharing the nonlocality with multiple observers in one side and found that up to six Charlies could share the standard tripartite nonlocality with a single Alice and a single Bob, and up to two Charlies could share the genuine tripartite nonlocality. In Ref. \cite{xi2023sharing}, the author found that an arbitrary number of Charlies could share the standard tripartite nonlocality with a single Alice and a single Bob. Furthermore, the bilateral sharing of nonlocality for two-qubit entangled states \cite{cheng2021limitations,zhu2022einstein} and the trilateral nonlocality sharing for three-qubit entangled states \cite{ren2022nonlocality} have also been studied. So far, significant progress has been made in the study of nonlocality sharing along this line of research\cite{hu2018observation,das2019facets,kumari2019sharing,foletto2020experimental,feng2020observation,zhang2022quantum,mahato2022sharing}.

Most of the studies on nonlocality sharing mentioned above have used unsharp measurements. Although projective measurement is the simplest form of measurement and is easily implemented in experiments, it is also the most destructive to quantum states. Entangled states become separable after projective measurements, thus limiting its application in nonlocality sharing. However, in recent work \cite{steffinlongo2022projective}, the authors demonstrated that if the Bobs choose to combine three projective measurement strategies with different probabilities, then two Bobs can share the nonlocality of the two-qubit entangled state with a single Alice. This opens up the study of nonlocality sharing using projective measurements. In Ref. \cite{zhang2024sharing} Zhang et al. investigated the scenario of bipartite high-dimensional pure states. Inspired by \cite{steffinlongo2022projective,zhang2024sharing}, this paper investigates the application of projective measurements in nonlocality sharing with a three-qubit entangled system. 

For the generalized GHZ state, we propose projective measurements and combine different measurement strategies to investigate nonlocality sharing. When considering unbiased measurement choices, where all possible measurement settings for each Charlie are uniformly distributed, two Charlies can share standard tripartite nonlocality with a single Alice and a single Bob. For different state parameters $\varphi$, we provide the feasible range for the double violation with respect to the combination probability $p$. However, unbiased measurement choices permit at most one Charlie to share genuine tripartite nonlocality with a single Alice and a single Bob. To overcome this limitation, we introduce the parameter $v$ to modify the unbiased measurement choices into biased measurement choices. This modification allows two Charlies to share genuine tripartite nonlocality with a single Alice and a single Bob. However, biased measurement choices do not increase the number of sequential observers sharing standard nonlocality. We also provide the feasible ranges for realizing nonlocality sharing with respect to the biased parameter $v$ and state parameter $\varphi$. Additionally, for two specific values of $v$, we present the feasible range of double violation concerning the combination probability $p$ and the state parameter $\varphi$.

\section{Defining tripartite nonlocality}\label{sec2}

In a  Bell scenario involving a three-qubit entangled state, there are three spatially separated parties, named Alice, Bob, and Charlie. They share a three-qubit entangled state, and perform the  measurements $A_x$, $B_y$, and $C_z$ on their subsystems, respectively, with outcomes $a$, $b$, and $c$ where $x, y, z \in\{0, 1\}$, $a, b, c \in \{+1, -1\}$. In this setup, the quantum correlations are described by the conditional probability $P(a,b,c |A_x,B_y,C_z)$. For all combinations of $x,y,z,a,b,c$, if the correlations $P(a,b,c |A_x,B_y,C_z)$ can be represented by a local hidden variable model,
\begin{equation}
    P(a,b,c|A_x,B_y,C_z)=\sum_\xi q(\xi)P_\xi(a|A_x)P_\xi(b|B_y,)P_\xi(c|C_z),
\end{equation}
where $q(\xi)$ is the probability distribution on the local hidden variable $\xi$, $0\le q(\xi)\le 1$ and $\sum_\xi q(\xi)=1$, then $\{P(a,b,c |A_x,B_y,C_z)\}$ is said to be fully local. If $\{P(a,b,c |A_x,B_y,C_z)\}$ is not fully local, it indicates standard tripartite nonlocality, which can be certified through the violation of the Mermin inequality \cite{mermin1990extreme}. This inequality takes the following form.
\begin{equation}
    M=\langle A_{1}B_{0}C_{0}\rangle+\langle A_{0}B_{1}C_{0}\rangle+\langle A_{0}B_{0}C_{1}\rangle-\langle A_{1}B_{1}C_{1}\rangle\leq2,
\end{equation}
where $\langle A_xB_yC_z\rangle=\sum_{abc}(abc)P(a,b,c|A_x,B_y,C_z)$.

\begin{figure}
    \centering
    \includegraphics[width=0.8\linewidth]{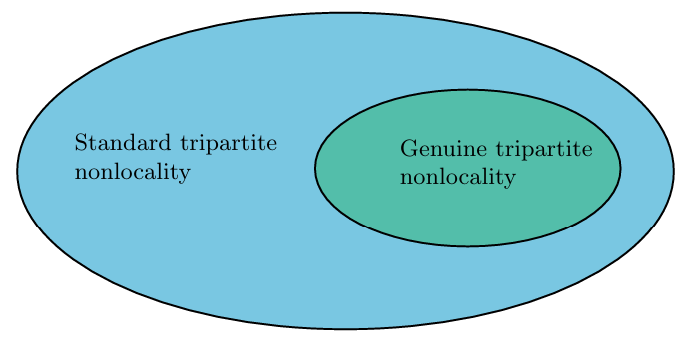}
        \caption{The relations among standard nonlocality and genuine nonlocality.}
        \label{fig1}
\end{figure}

In 1987, Svetlichny \cite{svetlichny1987distinguishing} introduced genuine tripartite nonlocality, which means that when the correlations cannot be described by the following local hidden variable model, 
\begin{equation}
    \begin{aligned}
        P(a,b,c|A_{x},B_{y},C_{z})& =\sum_{\xi}q(\xi)P_\xi(b,c|B_{y},C_{z})P_\xi(a|A_{x}) \\
        &+\sum_\mu q(\mu)P_\mu(a,c|A_x,C_z)P_\mu(b|B_y) \\
        &+\sum_\nu q(\nu)P_\nu(a,b|A_x,B_y)P_\nu(c|C_z),
    \end{aligned}
\end{equation}  
 where $0\le q(\xi),q(\mu),q(\nu)\le 1$, and $\sum_{\xi}q(\xi)+\sum_{\mu}q(\mu)+\sum_{\nu}q(\nu)=1$, it indicates genuine tripartite nonlocality. If a quantum correlation violates the Mermin inequality, it does not necessarily imply that the correlation exhibits genuine tripartite nonlocality. However, genuine tripartite nonlocality can be certified through the violation of the Svetlichny inequality \cite{svetlichny1987distinguishing}, which takes the following form
\begin{equation}
    \begin{aligned}
        S&=\langle A_0B_0C_1\rangle+\langle A_0B_1C_0\rangle+\langle A_1B_0C_0\rangle-\langle A_1B_1C_1\rangle\\&+\langle A_0B_1C_1\rangle+\langle A_1B_0C_1\rangle+\langle A_1B_1C_0\rangle-\langle A_0B_0C_0\rangle\leq4.    \end{aligned}
\end{equation}
According to the above definition, the relationship between standard tripartite nonlocality and genuine tripartite nonlocality can be described by Fig. \ref{fig1}.
\section{Sharing of standard tripartite nonlocality}\label{sec3}
Previous studies have achieved sequential sharing of standard tripartite nonlocality using unsharp measurements \cite{saha2019sharing,xi2023sharing}. In this section, we will explore whether projective measurements can enable multiple Charlies to share standard tripartite nonlocality with a single Alice and a single Bob.

\begin{figure}[htb]
    \centering
    \includegraphics[width=0.8\linewidth]{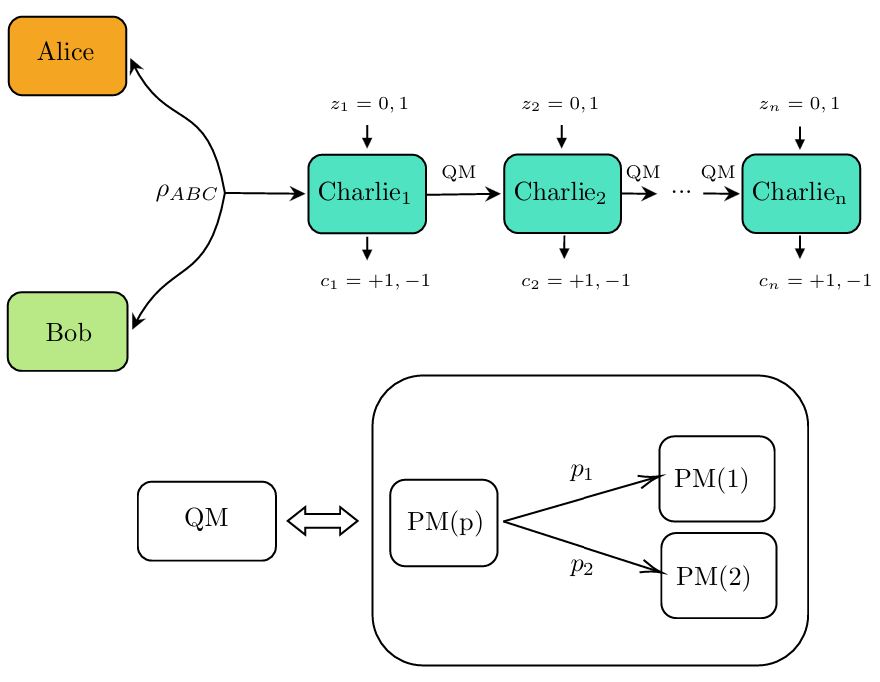}
    \caption{A quantum state $\rho_{ABC}$ is initially distributed between Alice, Bob, and $\mathrm{Charlie_1}$. After $\mathrm{Charlie_1}$ performs some kind of quantum measurements(QM) on his part and records the outcomes, he passes the postmeasurement quantum state to $\mathrm{Charlie_2}$, who then repeats the process. Where QM given by a random combination of several projective measurements (PMs) with different probabilities $p$. Before the experiment begins, all parties agree to share classical randomness $p =\{p_\lambda\}$ satisfying $\sum_{\lambda}p_\lambda=1$.}
    \label{fig2}
\end{figure}

 As shown in Fig. \ref{fig2}, three particles are prepared in the entangled source $\rho_{ABC}=|\text{GHZ}_{\varphi}\rangle \langle\text{GHZ}_{\varphi}|$, where $|\text{GHZ}_{\varphi}\rangle$ is the generalized three-qubit GHZ state.
\begin{equation}
    |\text{GHZ}_{\varphi}\rangle=\cos{\varphi }|000\rangle+\sin{\varphi }|111\rangle,\quad 0\le\varphi\le\pi/4.
    \label{eq5}
\end{equation}
  These three particles are spatially separated and shared between Alice, Bob, and multiple Charlies. Alice performs binary measurements on the first particle according to the input $x\in\{0,1\}$ and obtains the outcome $a\in\{+1,-1\}$. Bob performs binary measurements on the second particle according to the input $y\in\{0,1\}$ and obtains the outcome $b\in\{+1,-1\}$. $\text{Charlie}_k$ ($k = 1,...,n$) performs binary measurements on the third particle according to the input $z_k\in\{0,1\}$, obtains the outcome $c_k\in\{+1,-1\}$, and sends the postmeasurement state to the next Charlie, i.e., $\text{Charlie}_{k+1}$. Each Charlie can implement two different projective measurement strategies:
\textbf{PM(1)} ($\lambda=1$): Both measurements are projective measurements. \textbf{PM(2)} ($\lambda=2$): One measurement is a projective measurement and the other measurement is an identity measurement. 

It is crucial to determine the shared state $ \rho_{ABC}^{(k+1)} $ among Alice, Bob, and $\text{Charlie}_{k+1}$ after $\text{Charlie}_{k}$ performs measurements. Here, it is required that each Charlie performs measurements independent of the measurement choices and results of the preceding Charlies in the sequence, and we consider each observer’s input is equally probable. The postmeasurement states are determined by generalized von Neumann-Lüders transformation rule \cite{busch1986unsharp}:
\begin{equation}
    \rho_{ABC}^{(k+1,\lambda)}=\frac{1}{2}\sum_{c_k,z_k}\left(\mathbb{I}\otimes \mathbb{I}\otimes\sqrt{C_{c_{k}|z_{k}}^{(k,\lambda)}}\right)\rho_{ABC}^{(k,\lambda)}\left(\mathbb{I}\otimes \mathbb{I}\otimes\sqrt{C_{c_{k}|z_{k}}^{(k,\lambda)}}\right),
    \label{eq6}
\end{equation}
where $C_{c_{k}|z_{k}}^{(k,\lambda)}$ is the projective measurement, thus satisfying $\big(C_{c_{k}|z_{k}}^{(k,\lambda)}\big)^2=C_{c_{k}|z_{k}}^{(k,\lambda)}$.

From the previous section, we know that standard tripartite nonlocality can be certified through the violation of the Mermin inequality. Each pair Alice-Bob-$\text{Charlie}_k$ tests the Mermin inequality,
\begin{equation}
    M_k\equiv\sum_{\lambda=1}^2p_\lambda M_k^\lambda\leqslant2,
    \label{eq7}
\end{equation}
where 
\begin{equation}
    M_k^\lambda=\langle A_{1}B_{0}C_{0}^{(k,\lambda)}\rangle+\langle A_{0}B_{1}C_{0}^{(k,\lambda)}\rangle+\langle A_{0}B_{0}C_{1}^{(k,\lambda)}\rangle-\langle A_{1}B_{1}C_{1}^{(k,\lambda)}\rangle ,
\end{equation}
$\langle A_{x}B_{y}C_{z}^{(k,\lambda)}\rangle=\mathrm{Tr}\big[\rho_{ABC}^{(k,\lambda)}(A_{x}\otimes B_{y}\otimes C_{z}^{(k,\lambda)})\big]$ and $\{A_x,B_y,C_{z}^{(k,\lambda)}\}_{k=1,2,...,n}$ denote the observables of the respective parties conditioned on $\lambda$. Here, we consider the simplest scenario, namely, $n = 2$. 
For the generalized GHZ state (\ref{eq5}), we give the following measurment strategy for Alice, Bob, and $\text{Charlie}_k$.\\
Alice's observables are defined by:
\begin{equation}
    A_0=\sigma_x,\quad A_1=\sigma_y,
\end{equation}
Bob's observables are defined by:
\begin{equation}
    B_0=-\sigma_y,\quad B_1=\sigma_x.
\end{equation}
For $\text{Charlie}_k$, we separately analyze the two types of projective measurement strategies: Case(i)-(ii).\\
\textbf{Case(i):($\lambda=1$)} 
Both measurements of $\text{Charlie}_1$ are projective. The measurement settings are given by the following observables:
\begin{equation}
    C_{0|0}^{(1,1)}=\frac{\mathbb{I}+\sigma_x}2,\quad C_{0|1}^{(1,1)}=\frac{\mathbb{I}+\sigma_y}2.
\end{equation}
Using normalization and spectral decomposition theorem, we can obtain $C_{1|z}^{(1,1)}=\mathbb{I}-C_{0|z}^{(1,1)}$ and $C_{z}^{(1,1)}=C_{0|z}^{(1,1)}-C_{1|z}^{(1,1)}$ for $z=0,1$. Under this measurement strategy and the initial state $|\text{GHZ}_{\varphi}\rangle\langle\text{GHZ}_{\varphi}|$, we can calculate the Mermin inequality value for Alice, Bob, and $\text{Charlie}_1$ as follows:
\begin{equation}
    \begin{aligned}
         M_1^{\lambda=1}&=\mathrm{Tr}\big[\rho_{ABC}^{(1,1)}\big(A_{1}B_{0}C_{0}^{(1,1)}+A_{0}B_{1}C_{0}^{(1,1)}+A_{0}B_{0}C_{1}^{(1,1)}-A_{1}B_{1}C_{1}^{(1,1)}\big)\big]\\
         &=\mathrm{Tr}[\rho_{ABC}^{(1,1)}(-\sigma_{y}\otimes\sigma_{y}\otimes\sigma_{x}+\sigma_{x}\otimes\sigma_{x}\otimes\sigma_{x}-\sigma_{x}\otimes\sigma_{y}\otimes\sigma_{y}-\sigma_{y}\otimes\sigma_{x}\otimes\sigma_{y})]\\
          &=4\sin{2\varphi}.
    \end{aligned}
\end{equation}  
According to Eq. (\ref{eq6}), the state  shared among Alice, Bob, and $\text{Charlie}_2$ is given by
\begin{equation}
    \begin{aligned}
        \rho_{ABC}^{(2,1)}
        =\frac{1}{2}\rho_{ABC}^{(1,1)}+\frac{1}{4}(\mathbb{I}\otimes\mathbb{I}\otimes\sigma_{x})\rho_{ABC}^{(1,1)}(\mathbb{I}\otimes\mathbb{I}\otimes\sigma_{x})+\frac{1}{4}(\mathbb{I}\otimes\mathbb{I}\otimes\sigma_{y})\rho_{ABC}^{(1,1)}(\mathbb{I}\otimes\mathbb{I}\otimes\sigma_{y}).
    \end{aligned}
\end{equation}
Then taking $C_{z}^{(2,1)}=C_{z}^{(1,1)}$  for $z =0,1$, we can get
\begin{equation}
    \begin{aligned}
          M_2^{\lambda=1}&=\mathrm{Tr}\big[\rho_{ABC}^{(2,1)}\big(A_{1}B_{0}C_{0}^{(2,1)}+A_{0}B_{1}C_{0}^{(2,1)}+A_{0}B_{0}C_{1}^{(2,1)}-A_{1}B_{1}C_{1}^{(2,1)}\big)\big]\\
          &=2\sin{2\varphi}.
     \end{aligned}
\end{equation}
\textbf{Case(ii):($\lambda=2$)} 
One measurement of $\text{Charlie}_1$ is projective and the other is an identity measurement,
\begin{equation}
    C_{0|0}^{(1,2)}=\frac{\mathbb{I}+\sigma_x}2,\quad C_{0|1}^{(1,2)}=\mathbb{I}.
\end{equation}
Similarly, we can obtain $C_{1|z}^{(1,2)}=\mathbb{I}-C_{0|z}^{(1,2)}$, $C_{z}^{(1,2)}=C_{0|z}^{(1,2)}-C_{1|z}^{(1,2)}$ for $z=0,1$, and we can calculate the Mermin inequality value for Alice, Bob, and $\text{Charlie}_1$ as follows:
\begin{equation}
    \begin{aligned}
         M_1^{\lambda=2}&=\mathrm{Tr}\big[\rho_{ABC}^{(1,2)}\big(A_{1}B_{0}C_{0}^{(1,2)}+A_{0}B_{1}C_{0}^{(1,2)}+A_{0}B_{0}C_{1}^{(1,2)}-A_{1}B_{1}C_{1}^{(1,2)}\big)\big]\\
         &=\mathrm{Tr}[\rho_{ABC}^{(1,2)}(-\sigma_{y}\otimes\sigma_{y}\otimes\sigma_{x}+\sigma_{x}\otimes\sigma_{x}\otimes\sigma_{x} -\sigma_{x}\otimes\sigma_{y}\otimes\mathbb{I} -\sigma_{y}\otimes\sigma_{x}\otimes\mathbb{I})]\\
          &=2\sin{2\varphi}.
    \end{aligned}
\end{equation}  
The state  shared among Alice, Bob, and $\text{Charlie}_2$ is given by
\begin{equation}
    \begin{aligned}
        \rho_{ABC}^{(2,2)}
        =\frac{3}{4}\rho_{ABC}^{(1,2)}+\frac{1}{4}(\mathbb{I}\otimes\mathbb{I}\otimes\sigma_{x})\rho_{ABC}^{(1,2)}(\mathbb{I}\otimes\mathbb{I}\otimes\sigma_{x}).
    \end{aligned}
\end{equation}
Then taking $C_{0}^{(2,2)}=\sigma_x, C_{1}^{(2,2)}=\sigma_y$, we can get
\begin{equation}
    \begin{aligned}
          M_2^{\lambda=2}&=\mathrm{Tr}\big[\rho_{ABC}^{(2,2)}\big(A_{1}B_{0}C_{0}^{(2,2)}+A_{0}B_{1}C_{0}^{(2,2)}+A_{0}B_{0}C_{1}^{(2,2)}-A_{1}B_{1}C_{1}^{(2,2)}\big)\big]\\
          &=3\sin{2\varphi}.
     \end{aligned}
\end{equation}

Now we consider standard tripartite nonlocality based on the mixture of case(i) and case(ii). Let's assume the probability of choosing the first measurement is $p$, and the probability of choosing the second measurement is $1-p$. Then, from Eq. (\ref{eq7}), we have
\begin{equation}
    M_1\equiv p\cdot M_1^{\lambda=1}+(1-p)\cdot M_1^{\lambda=2}=p\cdot4\sin{2\varphi}+(1-p)\cdot2\sin{2\varphi}=(2p+2)\sin{2\varphi} ,
\end{equation}
and
\begin{equation}
    M_2\equiv p\cdot M_2^{\lambda=1}+(1-p)\cdot M_2^{\lambda=2}=p\cdot2\sin{2\varphi}+(1-p)\cdot3\sin{2\varphi}=(3-p)\sin{2\varphi} .
\end{equation}

\begin{figure}[htp]
   \centering
   \subfigure[]
        {\includegraphics[width=0.48\linewidth]{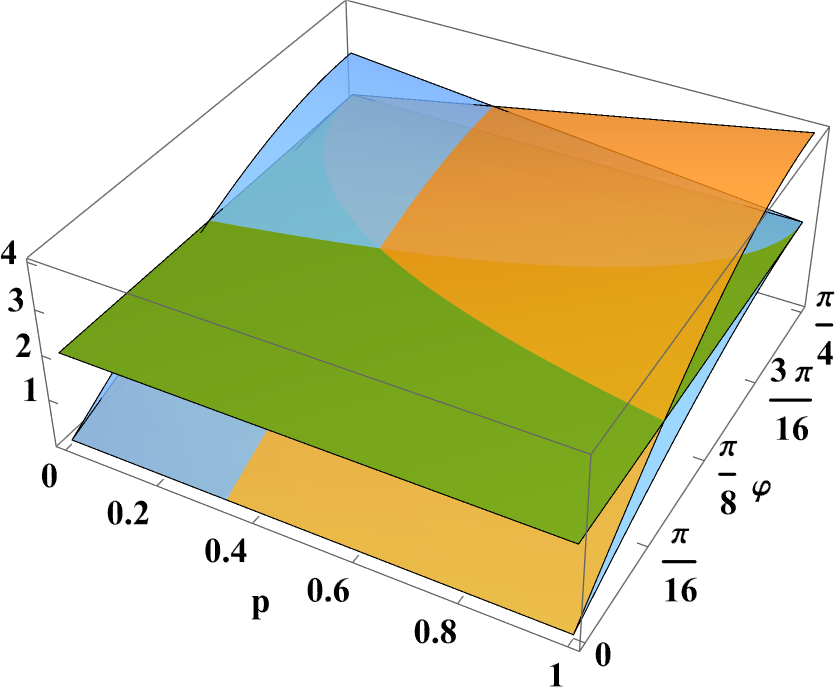}}  
    \subfigure[]     
        { \includegraphics[width=0.48\linewidth]{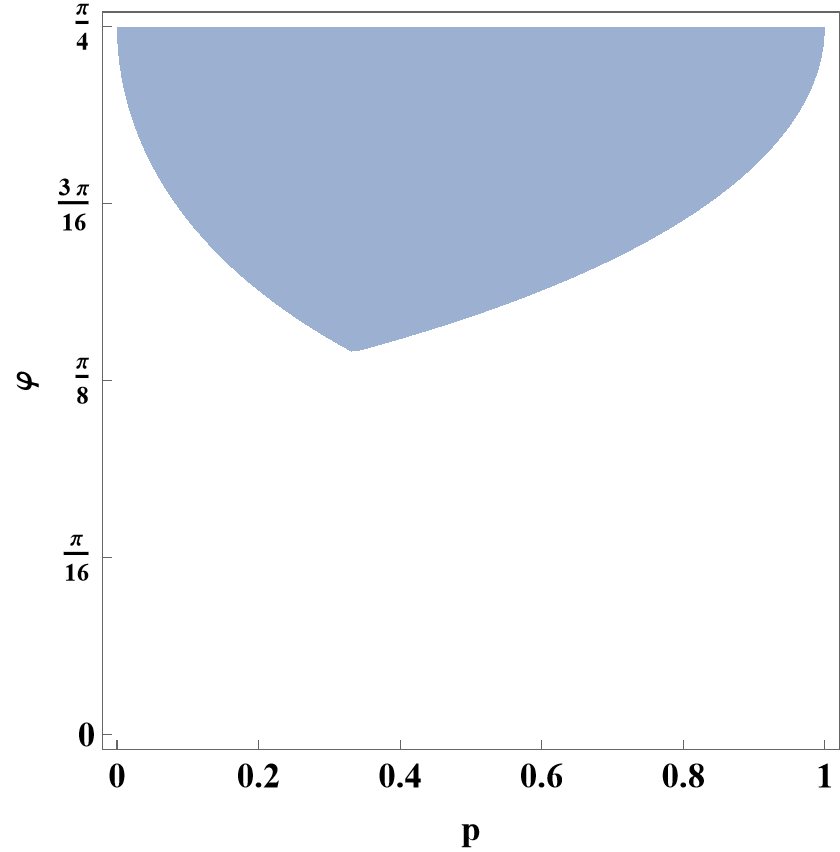}} 
    \caption{(a)\( M_1 \) (orange) and \( M_2 \) (blue) are violation surfaces parameterized by \( p \) and \( \varphi \), and the contour surface \( M = 2 \) (green) represents the local bound of the Mermin inequality. (b) The feasible range of parameters $p$ and $\varphi$ that satisfy the conditions.}
    \label{fig3}
   \end{figure}

Thus, the problem of nonlocality sharing can be transformed into determining whether we can find parameters $p$ and $\varphi$ such that both $M_1$ and $M_2$ are simultaneously greater than 2. In other words, the conditions $(2p+2)\sin{2\varphi}>2$ and $(3-p)\sin{2\varphi}>2$ must be satisfied. In Fig. \ref{fig3}, we plot the violations of the Mermin inequality \( M_1 \) and \( M_2 \) with respect to the parameters \( p \) and \( \phi \), and we can observe that there exist values of \( p \) and \( \varphi \) that satisfy the above two inequalities. And it can be observed that, as long as $\varphi\in(0.424,\pi/4]$, there exists a mixed strategy that allows two Charlies to share the standard nonlocality. For each fixed value of $\varphi$ within this range, the range of the parameter $p$ can be easily calculated, $\frac{1}{\sin{2\varphi}}-1< p< 3-\frac{2}{\sin{2\varphi}}$. 
\section{Sharing of  genuine tripartite nonlocality}
In Ref. \cite{saha2019sharing}, sequential sharing of the genuine tripartite nonlocality has been achieved using unsharp measurements. In this section, we will explore whether, for the generalized GHZ state, using only projective measurements can enable multiple Charlies to share the genuine tripartite nonlocality with a single Alice and a
single Bob. The measurement scenario is similar to that described in the Sect. \ref{sec3} and can be illustrated in Fig. \ref{fig2}.

From Sect. \ref{sec2}, we know that the genuine tripartite nonlocality can be certified through
the violation of the Svetlichny inequality. Each pair Alice-Bob-$\text{Charlie}_k$  tests the Svetlichny inequality,
\begin{equation}
    S_k\equiv\sum_{\lambda=1}^2p_\lambda S_k^\lambda\leqslant4,
    \label{21}
\end{equation}
where
\begin{equation}
    \begin{aligned}
        S_k^\lambda&=\langle A_0B_0C_{1}^{(k,\lambda)}\rangle+\langle A_0B_1C_0^{(k,\lambda)}\rangle+\langle A_1B_0C_0^{(k,\lambda)}\rangle-\langle A_1B_1C_1^{(k,\lambda)}\rangle\\&+\langle A_0B_1C_1^{(k,\lambda)}\rangle+\langle A_1B_0C_1^{(k,\lambda)}\rangle+\langle A_1B_1C_0^{(k,\lambda)}\rangle-\langle A_0B_0C_0^{(k,\lambda)}\rangle.
    \end{aligned}
\end{equation}

Here, we also consider the simplest scenario, and give the following measurement strategy for Alice, Bob, and $\text{Charlie}_k$.\\
Alice’s observables as follows:
\begin{equation}
    A_0=\sigma_x,\quad A_1=\sigma_y,
\end{equation}
Bob’s observables as follows:
\begin{equation}
    B_0=\frac{1}{\sqrt{2}}(\sigma_x-\sigma_y),\quad B_1=\frac{1}{\sqrt{2}}(\sigma_x+\sigma_y).
\end{equation}
Next, the measurement strategies of $\text{Charlie}_k$ are divided into Case(i) and Case(ii).\\
\textbf{Case(i):($\lambda=1$)}Both measurements of $\text{Charlie}_1$ are projective,
\begin{equation}
    C_{0|0}^{(1,1)}=\frac{\mathbb{I}-\sigma_y}2,\quad C_{0|1}^{(1,1)}=\frac{\mathbb{I}+\sigma_x}2.
\end{equation}
We can obtain $C_{1|z}^{(1,1)}=\mathbb{I}-C_{0|z}^{(1,1)}$, $C_{z}^{(1,1)}=C_{0|z}^{(1,1)}-C_{1|z}^{(1,1)}$ for $z=0,1$, and it is not difficult to calculate that the Svetlichny inequality value for Alice, Bob, and $\text{Charlie}_1$ is $S_1^{\lambda=1}=4\sqrt{2}\sin{2\varphi}$.\\
Using Eq. (\ref{eq6}) we obtain
 \begin{equation}
        \rho_{ABC}^{(2,1)}=\frac{1}{2}\rho_{ABC}^{(1,1)}+\frac{1}{4}(\mathbb{I}\otimes\mathbb{I}\otimes\sigma_{x})\rho_{ABC}^{(1,1)}(\mathbb{I}\otimes\mathbb{I}\otimes\sigma_{x})+\frac{1}{4}(\mathbb{I}\otimes\mathbb{I}\otimes\sigma_{y})\rho_{ABC}^{(1,1)}(\mathbb{I}\otimes\mathbb{I}\otimes\sigma_{y}).
\end{equation}
Then taking $C_{z}^{(2,1)}=C_{z}^{(1,1)}$ for $z=0,1$ ,we can get $S_2^{\lambda=1}=2\sqrt{2}\sin{2\varphi}$.\\
\textbf{Case(ii):($\lambda=2$)} One measurement of $\text{Charlie}_1$ is projective and the other is an identity measurement, and their measurement settings are given by the following observables:
\begin{equation}
    C_{0|0}^{(1,2)}=\mathbb{I},\quad C_{0|1}^{(1,2)}=\frac{\mathbb{I}+\sigma_x}2.
    \label{27}
\end{equation}
We can obtain $C_{1|z}^{(1,2)}=\mathbb{I}-C_{0|z}^{(1,2)}$, $C_{z}^{(1,2)}=C_{0|z}^{(1,2)}-C_{1|z}^{(1,2)}$ for $z=0,1$. If, similar to Sect. \ref{sec3}, we select the two measurements in Eq. (\ref{27}) with equal probability, we find that at most one Charlie can share the genuine tripartite nonlocality with a single Alice and a single Bob. Therefore, we will use biased measurement choices here. Let's assume $\text{Charlie}_1$ selects measurement $C_{0}^{(1,2)}$ with the probability of $v$, and measurement $C_{1}^{(1,2)}$ with the probability of $1-v$, where $v \in (0,1)$. We can
calculate that the Svetlichny inequality value for Alice, Bob, and $\text{Charlie}_1$ is $S_1^{\lambda=2}=2\sqrt{2}\sin{2\varphi}$. 
The state shared among Alice, Bob, and $\text{Charlie}_2$ is given by
\begin{equation}
    \begin{aligned}
        \rho_{ABC}^{(2,2)}
        &=(1-v)\left(\mathbb{I}\otimes\mathbb{I}\otimes\frac{\mathbb{I}+\sigma_{x}}{2}\rho_{ABC}^{(1,2)}\mathbb{I}\otimes\mathbb{I}\otimes\frac{\mathbb{I}+\sigma_{x}}{2}+\mathbb{I}\otimes\mathbb{I}\otimes\frac{\mathbb{I}-\sigma_{x}}{2}\rho_{ABC}^{(1,2)}\mathbb{I}\otimes\mathbb{I}\otimes\frac{\mathbb{I}-\sigma_{x}}{2}\right)\\
        &\quad+v\cdot\mathbb{I}\otimes\mathbb{I}\otimes\mathbb{I}\ \rho_{ABC}^{(1,2)}\ \mathbb{I}\otimes\mathbb{I}\otimes\mathbb{I}\\
        &=\frac{1+v}{2}\rho_{ABC}^{(1,2)}+\frac{1-v}{2}(\mathbb{I}\otimes\mathbb{I}\otimes\sigma_{x})\rho_{ABC}^{(1,2)}(\mathbb{I}\otimes\mathbb{I}\otimes\sigma_{x}).
    \end{aligned}
\end{equation}
Then taking $C_{0}^{(2,2)}=-\sigma_y, C_{1}^{(2,2)}=\sigma_x$, we can get $S_2^{\lambda=2}=2\sqrt{2}(1+v)\sin{2\varphi}$.\\
Similar to the Sect. \ref{sec3}, we now consider the mixture of case(i) and case(ii). Let's assume the probability of choosing the first measurement is $p$, and the probability of choosing the second measurement is $1-p$. From Eq. (\ref{21}), we have
\begin{equation}
\begin{aligned}
    S_1
    &\equiv p\cdot S_1^{\lambda=1}+(1-p)\cdot S_1^{\lambda=2}=p\cdot4\sqrt{2}\sin{2\varphi}+(1-p)\cdot2\sqrt{2}\sin{2\varphi}\\
    &=2\sqrt{2}(p+1)\sin{2\varphi},
\end{aligned}   
\end{equation}
and
\begin{equation}
\begin{aligned}
    S_2
    &\equiv p\cdot S_2^{\lambda=1}+(1-p)\cdot S_2^{\lambda=2}=p\cdot2\sqrt{2}\sin{2\varphi}+(1-p)\cdot2\sqrt{2}(1+v)\sin{2\varphi}\\
    &=2\sqrt{2}\big[1+v(1-p)\big]\sin{2\varphi}.
\end{aligned}
\end{equation}
\begin{figure}[htp]
    \centering
    \includegraphics[width=0.6\linewidth]{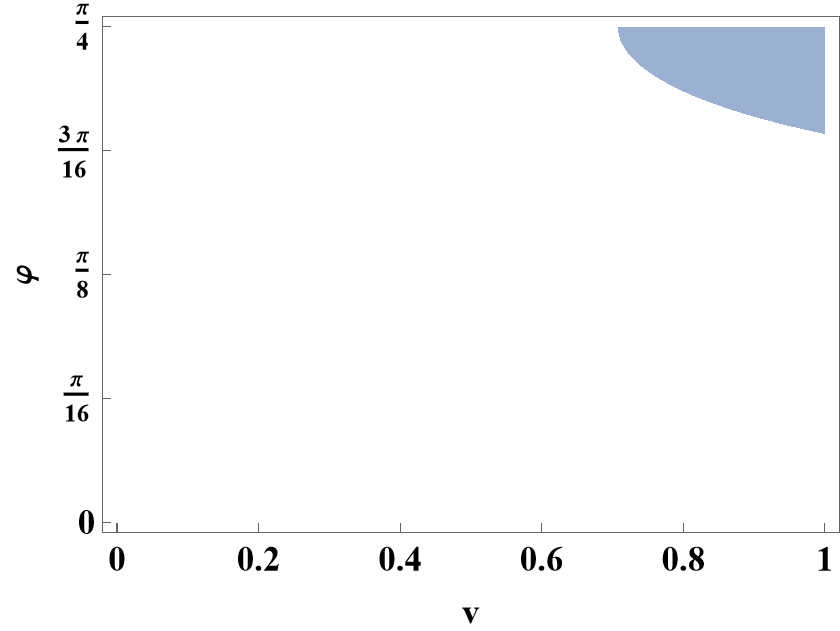}
    \caption{When the parameters $ v $ and $ \varphi $ are within the blue region, there exists a mixed strategy such that both $ S_1 $ and $ S_2 $ are greater than 4.}
    \label{fig4}
\end{figure}

\begin{figure}[htp]
   	\centering
   	\subfigure[$v=0.8$]{\label{5a} 
   		\includegraphics[width=0.48\linewidth]{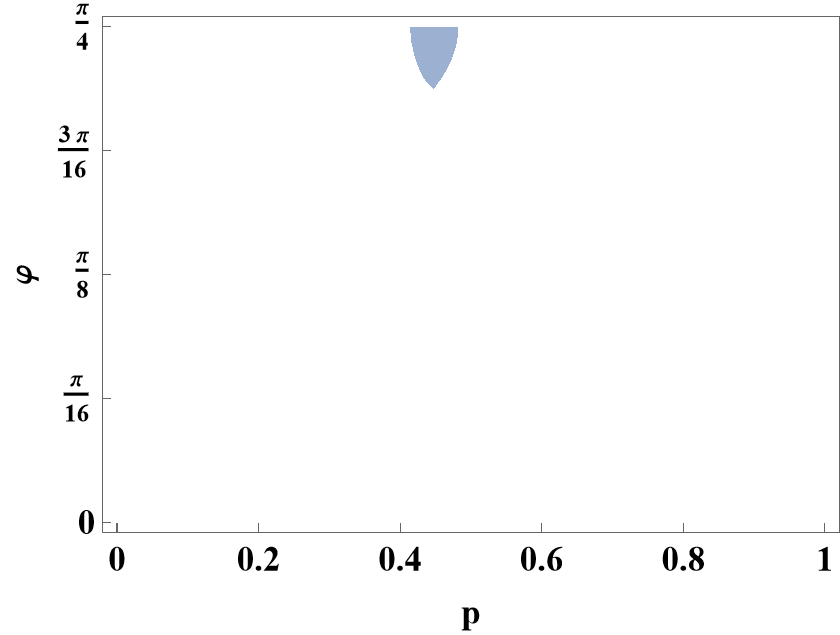}}  
    \subfigure[$v=0.9$]{\label{5b} 
      \includegraphics[width=0.48\linewidth]{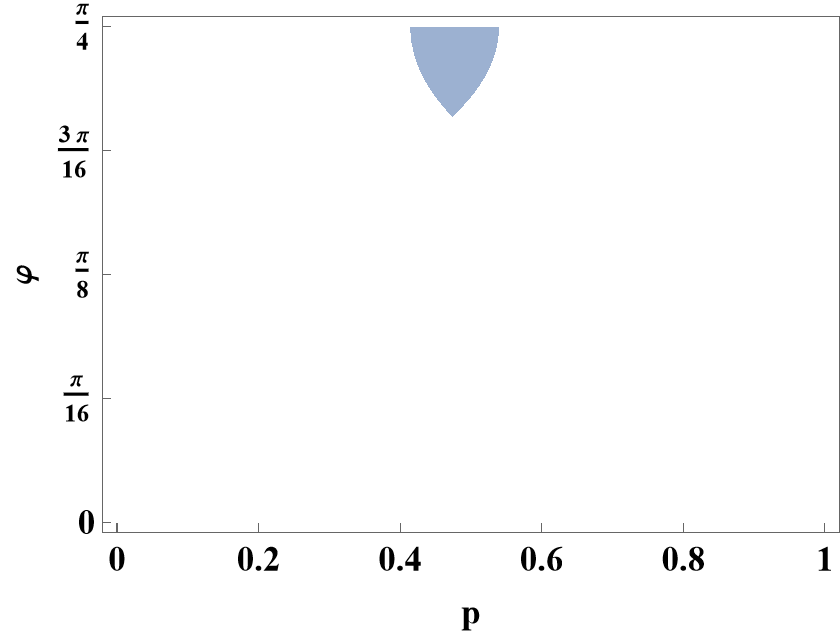}} 
     
    \caption{(a) and (b) show the feasible ranges of \( \varphi \) and \( p \) for the double violation when \( v = 0.8 \) and \( v = 0.9 \), respectively.}
    \label{fig5}
   \end{figure}

Now, we need to investigate whether we can find parameters $v$, $p$, and $\varphi$ such that both  $S_1$  and $ S_2$ are simultaneously greater than 4. In other words, it needs to satisfy
\begin{equation}
    2\sqrt{2}(p+1)\sin{2\varphi}>4,\ \text{and}\ 2\sqrt{2}\big[1+v(1-p)\big]\sin{2\varphi}>4,
    \label{eq31}
\end{equation}
 which implies 
 \begin{equation}
     0\le\frac{\sqrt{2}}{\sin{2\varphi}}-1< p< 1+\frac{1}{v}-\frac{\sqrt{2}}{v\sin{2\varphi}}\le1.
 \end{equation}
 In Fig. \ref{fig4}, we can observe that there exist \( v \) and \( \varphi \) such that both inequalities in Eq. (\ref{eq31}) hold. It is found that when the range of \( v \) is \( (0.7071,1) \), with some state  parameters $\varphi$ there exists a combination of measurements such that both \( S_1 \) and \( S_2 \) are simultaneously greater than 4. This indicates that unbiased measurements cannot achieve double violations. For example, when selecting $ v = 0.8 $ and $ v = 0.9 $, we obtain Fig. \ref{fig5}. It can be observed that as \( v \) increases, the feasible range for double violation with respect to \( \varphi \) and \( p \) also expands. In Fig. \ref{5a}, we observe that as long as \( \varphi \in (0.683, \pi/4] \), there exists a mixed strategy such that \( S_1 \) and \( S_2 \) are both greater than 4. Specifically, when \( \varphi = \pi/4 \), \( p \in (\sqrt{2} - 1, \frac{9 - 5\sqrt{2}}{4}) \approx (0.4143, 0.4822) \), where \( S_1 \) and \( S_2 \) are simultaneously greater than 4. Similarly, in Fig. \ref{5b}, we find that when \( \varphi \in (0.643, \pi/4] \), there exists a mixed strategy, and when \( \varphi = \pi/4 \), the feasible range for \( p \) is (\(\sqrt{2} - 1, \frac{19 - 10\sqrt{2}}{9}) \approx (0.4143, 0.5397) \).

 Finally, it can be seen that using biased measurement choices can increase the number of sequential observers sharing the genuine tripartite nonlocality. However, through calculations, we find that even with biased measurement choices, at most two Charlies can share the standard nonlocality with a single Alice and a single Bob.

\section{Conclusion}\label{conclude}

We have demonstrated that three-qubit nonlocality sharing can be achieved solely through projective measurements when the parties share classical randomness. Specifically, we found that unbiased measurement choices enable two Charlies to share the standard tripartite nonlocality with a single Alice and a single Bob. Furthermore, biased measurement choices allow two Charlies to share the genuine tripartite nonlocality with a single Alice and a single Bob. Additionally, we investigated the sharing of tripartite nonlocality among bilateral and trilateral scenarios. However, we found that with the measurement settings in this paper, it is not possible to achieve the sharing of standard tripartite nonlocality and genuine nonlocality among more than one sequential observer. Our results suggest that many other sequential quantum information protocols, such as steering \cite{silva2015multiple}, entanglement witnessing \cite{bera2018witnessing,pandit2022recycled}, and contextuality \cite{anwer2021noise}, can also be implemented based on projective measurements.

The current work raises some interesting questions: (1) In \cite{xi2023sharing}, it was proven that using unsharp measurements, any number of Charlies can share standard nonlocality by violating the Mermin inequality with a single Alice and a single Bob. It is still unknown whether more than two sequential violations can be achieved using projective measurements. (2) Whether there exist some state and measurement strategies such that tripartite nonlocal correlations can be shared among a single Alice—multiple Bobs—multiple Charlies, and multiple Alices—multiple Bobs—multiple Charlies.
\section{Acknowledgments}
This work is supported by the National Natural Science Foundation of China (Grants No. 62171056, and No. 62220106012), and the Henan Key Laboratory of Network Cryptography Technology (Grants No. LNCT2022-A03).

 \bibliographystyle{unsrt}
 \bibliography{sample}

\end{document}